\documentclass{elsarticle}

\usepackage{hyperref}

\journal{Journal of Computational Science}

\usepackage{amssymb}
\usepackage{amsmath}
\usepackage{amsfonts}
\def\AA{{\bf A}}
\def\DD{{\bf D}}
\def\II{{\bf I}}
\def\JJ{{\bf J}}
\def\UU{{\bf U}}
\def\KK{{\bf K}}
\def\ff{{\bf f}}

\def\rr{{\bf r}}
\def\mm{{\bf m}}
\def\ww{{\bf w}}
\def\hh{{\bf h}}
\def\PP{{\bf P}}
\def\II{{\bf I}}
\def\CC{{\bf C}}
\def\VV{{\bf V}}
\def\XX{{\bf X}}
\def\WW{{\bf W}}
\def\SS{{\bf \Sigma}}

\def\llambda{{\boldsymbol \lambda}}
\DeclareMathOperator*{\diag}{diag}

\begin{document}

\begin{frontmatter}

\title{Reduced Order Models for Pricing European and American Options under Stochastic
Volatility and Jump-Diffusion Models}

\author[address1]{Maciej Balajewicz\corref{mycorrespondingauthor}}
\cortext[mycorrespondingauthor]{Corresponding author}
\ead{mbalajew@illinois.edu}

\author[address2,address3]{Jari Toivanen}
\ead{toivanen@stanford.edu, jari.toivanen@jyu.fi}

\address[address1]{University of Illinois at Urbana-Champaign,
  Champaign, Illinois, U.S.A.}
\address[address2]{Stanford University,
  Stanford, California, U.S.A.}
\address[address3]{University of Jyv\"askyl\"a,
  Jyv\"askyl\"a, Finland}



\begin{abstract}
European options can be priced by solving parabolic partial(-integro)
differential equations under stochastic volatility and jump-diffusion models
like Heston, Merton, and Bates models. American option prices can be obtained by
solving linear complementary problems (LCPs) with the same operators. A finite
difference discretization leads to a so-called full order model (FOM). Reduced
order models (ROMs) are derived employing proper orthogonal decomposition (POD).
The early exercise constraint of American options is enforced by a penalty on
subset of grid points. The presented numerical experiments demonstrate that
pricing with ROMs can be orders of magnitude faster within a given model
parameter variation range.
\end{abstract}


\end{frontmatter}


\section{Introduction}
\label{sect:introduction}

European options can be exercised only at expiry while American options
can be exercised anytime until expiry. Due to this additional flexibility
the American options can be more valuable. In order to avoid arbitrage
the price must be always at least the same as the final payoff function.
A put option gives the right to sell the underlying asset for a specified
strike price while a call option gives the right to buy the asset for
a strike price.
The seminal paper \cite{Black73} by Black and Scholes employs a geometrical
Brownian motion with a constant volatility as a model for the price of
the underlying asset. The market prices of options show that the volatility
varies depending on the strike price and expiry of option. Several more
generic models for the asset prices have been developed which are more
consistent with market prices. Merton proposed adding log-normally
distributed jumps to this model \cite{Merton76}. Heston \cite{Heston93} made
the volatility to be a mean reverting stochastic process. Bates \cite{Bates96}
combined the Heston stochastic volatility model and Merton jump-diffusion
model.

There are many methods for pricing options. The Monte Carlo method simulate
asset price paths to compute the option price. This is an intuitive and
flexible method, but it can be slow when high precision is require and
it is more complicated and less efficient for American options. Instead
in this paper, the pricing is based on partial(-integro) differential
equation (P(I)DE) formulations. Another approach is based on numerical
integration techniques. One benefit of these formulations is that for
many options they can provide a highly accurate price much faster than the
Monte Carlo method. Here the European options are priced by solving a
P(I)DE and the American options by solving an LCP with the same operator.
These operators are two-dimensional with a stochastic volatility and
one-dimensional otherwise. The potential integral part of the model
results from the jumps.

The most common way to discretize the differential operators is the finite
difference method. For European and American options the discrization leads
to a system of linear equations and an LCP, respectively, at each time step.
Under stochastic volatility models efficient PDE based methods for American
options have been considered in \cite{Clarke99,Haentjens15,Ikonen08,Zvan98},
for example. A penalty approximation is employed for the resulting LCPs in
\cite{Zvan98} and an operator splitting method in \cite{Haentjens15,Ikonen08}.
An alternating direction implicit (ADI) method is used in \cite{Haentjens15}
while iterative methods are used for resulting linear systems in
\cite{Clarke99,Ikonen08,Zvan98}. Under jump-diffusion models PIDE methods
lead to a system with a full matrix at each time step and their efficient
solution for American options has been considered in
\cite{dHalluin04,Kwon11a,Salmi11,Salmi12,Salmi14a}, for example.
A penalty method together with FFT based fast method for evaluating the jump
integral was used in \cite{dHalluin04}. An iterative method was proposed
for LCPs with full matrices in \cite{Salmi11}. An implicit-explicit (IMEX)
method was proposed in \cite{Kwon11a} to treat the integral term explicitly
and the same approach was studied in \cite{Salmi14a}.
The generalizations of the above methods for the combined Bates model
have been developed and studied in
\cite{Toivanen10,Ballestra10,Salmi13,Salmi14b,vonSydow15}.

Unfortunately, such high-fidelity simulations are still too expensive for many
practical applications and reduced order modeling (ROM) is a promising tool for
significantly alleviating computational costs~\cite{antoulas2005,benner2015}.
Most existing ROM approaches are
based on projection. In projection-based reduced order modeling the state
variables are approximated in a low-dimensional subspace. Bases for this
subspace are typically constructed by Proper Orthogonal Decomposition
(POD)~\cite{sirovich87} of a set of high-fidelity solution snapshots. 
While many approaches have already been developed for the efficient reduction of
linear computational models 
three main strategies have been explored so far for efficiently reducing
nonlinear computational models.  The first one is based on linearization
techniques~\cite{Rewienski_LAA_2006,Gu_IEEE_2008}.  The second one is based on
the notion of
precomputations~\cite{Barbic_ACM_2005,Balajewicz_JFM_2013,
Balajewicz_JCP_2016,Balajewicz_ND_2012,Cordier_EF_2013},
but is limited to polynomial nonlinearities. The third strategy relies on the
concept of hyper-reduction --- that is, the approximation of the reduced
operators underlying a nonlinear reduced-order model (ROM) by a scalable
numerical technique based on a reduced computational
domain~\cite{Ryckelynck_JCP_2005,An_TOG_2008,chaturantabut2010,
	Carlberg_IJNME_2011,Amsallem_IJNME_2012,Farhat_IJNME_2014,Amsallem_SMO_2014,Farhat_IJNME_2015}. 

In the case
where the governing equations include a constraint equation it is often
beneficial to construct a basis that satisfies these constraints a
priori~\cite{Burkovska15}. For example, in the case of non-negativity
constraints, a non-negative basis can be constructed via non-negative matrix
factorization (NNMF)~\cite{Balajewicz15}. This approach was employed for
option pricing in \cite{Balajewicz16}. 

For pricing European options ROMs have been developed in \cite{Cont11,Sachs13}.
Only recently ROMs have been applied for pricing American options in
\cite{Burkovska15,Haasdonk13}.
A common problem associated with option pricing is the calibration of
model parameters to correspond to the market prices of options. This
is typically formulated as a least squares -type optimization problem.
The calibration is computationally expensive as it requires pricing
a large number of options with varying parameters. The use of ROMs
to reduce this computational cost has been studied in
\cite{Pironneau09,Sachs10,Sachs14}.

The main contribution of the present work is the development of a cheap and
accurate hyper-reduction approach for the early exercise constraint of American
options. Our proposed approach is based on the fact that accurate price
predictions do not necessarily require accurate approximations of the Lagrange
multipliers. This has been observed in practice for the reduction of structural
contact problems~\cite{Balajewicz15}. Our numerical
experiments summarized in this paper suggest that using the binary matrix
as the basis for the Lagrange multipliers performs remarkably well for all
reproductive and predictive simulations considered. This approach is simpler,
faster, and comparable in accuracy to previous approaches based on the
NNMF~\cite{Balajewicz16}.

This paper is organized as follows. In Section~\ref{sect:FOMs}, the full order
models considered in this work are overviewed. In Section~\ref{sect:ROMs} the
proposed new ROM approach is laid out. In Section~\ref{sect:Numerics}
the proposed approach is applied to several problems. Finally in
Section~\ref{sec:conclusions}, conclusions are offered and prospects for future work
are summarized. 

\section{Full Order Models}
\label{sect:FOMs}
Merton \cite{Merton76} proposed the price $s \ge 0$ of an underlying asset to
follow the stochastic differential equation
\begin{equation}
ds = (g - \mu \xi) s dt + \sigma_s s dw_s + s dJ,
\end{equation}
where $t$ is the time, $g$ is the growth rate of the asset price, $\sigma_s$
is its volatility, $w_s$ is a Wiener process, and $J$ is a compound Poisson
process with the jump intensity $\mu$ and the log-normal jump distribution
\begin{equation}
p(y) = \frac{1}{y \delta \sqrt{2\pi}} \exp
\left( - \frac{(\log y - \gamma)^2}{2 \delta^2} \right).
\end{equation}
The relative expected jump is
$\xi = \exp \left( \gamma + \tfrac{1}{2} \delta^2 \right) - 1$.
The Black--Scholes model is obtained by setting the jump intensity
$\mu$ to zero.
Under the Merton model the price $u(s,\tau)$ of a European option
can be obtained by solving the one-dimensional PIDE
\begin{equation}\label{Merton}
\frac{\partial u}{\partial \tau} =
\frac{1}{2} \sigma_s^2 s^2 \frac{\partial^2 u}{\partial s^2}
+ (r  - \mu \xi) s \frac{\partial u}{\partial s}
- (r + \mu) u + \mu \int_{0}^\infty u(sy,\tau) p(y) dy
=: L^M u,
\end{equation}
where $\tau = T - t$ is the time until expiry, $T$ is the expiry time, 
$r$ is the interest rate.

Bates \cite{Bates96} proposed the price $s$ and its instantaneous
variance $v \ge 0$ to follow the stochastic differential equations
\begin{equation}
\begin{split}
ds &= (g - \mu \xi) s dt + \sqrt{v} s dw_s + s dJ \\
dv &= \kappa (\theta - v) dt + \sigma_v \sqrt{v} dw_v, \\
\end{split}
\end{equation}
where $\theta$ is the mean level of $v$, $\kappa$ is the rate of mean reversion,
$\sigma_v$ is the volatility of $\sqrt{v}$, and $w_v$ is a Wiener process.
The Wiener prosesses $w_s$ and $w_v$ have the correlation $\rho$.
Under the Bates model the price $u(s,v,\tau)$ of a European option
can be obtained by solving the two-dimensional PIDE
\begin{equation}\label{Bates}
\begin{split}
\frac{\partial u}{\partial \tau} &{} =
\frac{1}{2} v s^2 \frac{\partial^2 u}{\partial s^2}
+ \rho \sigma_v v s \frac{\partial^2 u}{\partial s \partial v}
+ \frac{1}{2} \sigma_v^2 v \frac{\partial^2 u}{\partial v^2}
+ (r - \mu \xi) s \frac{\partial u}{\partial s}
+ \kappa ( \theta - v ) \frac{\partial u}{\partial v} \\
&{} - (r + \mu) u
+ \mu \int_{0}^\infty u(sy,v,\tau) p(y) dy
=: L^B u,
\end{split}
\end{equation}
The Heston model is obtained by setting the jump intensity $\mu$
to zero.

In the following, put options are considered. Their price at the expiry
is given by the pay-off function $g(s) = \max\{K - s,\, 0\}$.
As the equations are solved backward in time, this leads to the
initial condition
\begin{equation}\label{initial}
u(s,0) = g(s)
\qquad\text{and}\qquad
u(s,v,0) = g(s)
\end{equation}
for one-dimensional and two-dimensional models, respectively.

For computing an approximate solution the infinite domain is truncated
at $s = s_{\max}$ and $v = v_{\max}$, where $s_{\max}$ and $v_{\max}$
are sufficiently large so that the error due to truncation is negligible.
The price $u$ of a European put option satisfies the Dirichlet boundary
conditions
\begin{equation}
u = K e^{-r \tau}\;\;\text{at}\;\; s = 0
\qquad\text{and}\qquad
u = 0\;\;\text{at}\;\; s = s_{\max}.
\end{equation}
For a non-negative interest rate $r \ge 0$,
the price $u$ of an American put option satisfies the Dirichlet boundary
conditions
\begin{equation}
u = K\;\;\text{at}\;\; s = 0
\qquad\text{and}\qquad
u = 0\;\;\text{at}\;\; s = s_{\max}.
\end{equation}
Under the stochastic volatility models, the Neumann boundary condition
$\tfrac{\partial u}{\partial v} = 0$ is posed at $v = v_{\max}$.
The second derivatives in \eqref{Bates} vanish on the boundary $v = 0$.
It is shown in \cite{Ekstrom10} that this degenerated form defines
appropriate boundary condition at $v = 0$.

Due to early exercise possibility the price $u$ of an American option
satisfies the LCP
\begin{equation}\label{American}
\tfrac{\partial u}{\partial \tau} - L u = \lambda, \quad
u \ge g, \quad
\lambda \ge 0, \quad
\lambda (u - g) = 0,
\end{equation}
where the operator $L$ is either $L^M$ or $L^B$ depending on the model
and $\lambda$ is a Lagrange multiplier; see \cite{Ito09}, for example.

For an easier numerical solution, the P(I)DE for European options and
the LCP for American options are reformulated for $w$ which
satisfies the homogeneous Dirichlet boundary condition $w = 0$
at $s = 0$. Furthermore, $w$ for American options is chosen so
that it satisfies the positivity constraint $w \ge 0$ instead
of the more complicated constraint $u \ge g$.
For European options $w$ is chosen to be $w = u - e^{-r \tau} g$
while for American options it is chosen to be $w = u - g$.

For European options the choice $w = u - e^{-r \tau} g$ leads to the P(I)DE
\begin{equation}\label{formwe}
\tfrac{\partial w}{\partial \tau} - L w = e^{-r \tau} (L + r) g.
\end{equation}
For American options the choice $w = u - g$ leads to the LCP
\begin{equation}\label{formw}
\tfrac{\partial w}{\partial \tau} - L w = \lambda + L g, \quad
w \ge 0, \quad
\lambda \ge 0, \quad
\lambda w = 0.
\end{equation}

For American options a quadratic penalty formulation is obtained
by choosing the Lagrange multiplier to be
\begin{equation}\label{lambdadef}
\lambda = -\tfrac{1}{\varepsilon} \max \left\{-w,0\right\} w.
\end{equation}
This leads to the nonlinear P(I)DE
\begin{equation}
\tfrac{\partial w}{\partial \tau} - L w
+ \tfrac{1}{\varepsilon} \max \left\{-w,\, 0\right\} w = L g.
\end{equation}

For the finite difference discretization, a grid is defined by $s_i$,
$i = 0, 1, 2, \ldots, N_s$, for the interval $[0, s_{\max}]$.
The spatial partial derivatives with respect to $s$ are discretized using
central finite difference
\begin{equation}
\frac{\partial w}{\partial s} (s_i) \approx
\tfrac{1}{\Delta s_{i-1} + \Delta s_i} \left[
- \tfrac{\Delta s_i}{\Delta s_{i-1}} w_{i-1}
+ \left( \tfrac{\Delta s_i}{\Delta s_{i-1}}
- \tfrac{\Delta s_{i-1}}{\Delta s_i} \right) w_i
+ \tfrac{\Delta s_{i-1}}{\Delta s_i} w_{i+1}
\right]
\end{equation}
and
\begin{equation}
\frac{\partial^2 w}{\partial s^2} (s_i) \approx
\tfrac{2}{\Delta s_{i-1} + \Delta s_i} \left[
\tfrac{1}{\Delta s_{i-1}} w_{i-1}
- \left( \tfrac{1}{\Delta s_{i-1}} + \tfrac{1}{\Delta s_i} \right) w_i
+ \tfrac{1}{\Delta s_i} w_{i+1}
\right],
\end{equation}
where $\Delta s_i = s_{i+1} - s_i$.
Similarly for the interval $[0, v_{\max}]$, a grid is defined
$v_j$, $j = 0, 1, 2, \ldots, N_v$.
The spatial partial derivatives with respect to $v$ are discretized
using the above central finite differences.
A nine-point finite difference stencil for
$\tfrac{\partial^2 w}{\partial s \partial v}$ is obtained by employing
the central finite differences in both directions.
While this approximation can be unstable with high correlations
$\rho$ it is stable for numerical experiments presented in
Section \ref{sect:Numerics}.
On the boundary $v = 0$, a one-sided finite difference approximation is
used for $\tfrac{\partial w}{\partial v}$.
The integrals can be discretized using a second-order accurate
quadrature formula. Here the linear interpolation is used for $w$ between
grid points and exact integration; see \cite{Salmi11}, for details.
Under the Merton model the discretization of the integral leads to a full
matrix while under the Bates model it leads to full diagonal blocks.

Under models without jumps the time discretization is performed by
taking the first time steps using the implicit Euler method and after
using the second-order accurate BDF2 method. Under jump models the
integral is treated explicitly. In the first time step using the explicit
Euler method and in the following time steps using the linear extrapolation
based on the two previous time steps. This IMEX-BDF2 method is described
in \cite{Salmi14a}. With the explicit treatment of the integral it is not
necessary to solve systems with dense matrices.
At the time $(k+1) \Delta \tau$, the grid point values contained in
the vector $\ww^{k+1}$ are obtained by solving the system
\begin{equation}\label{BDF2stepe}
\left( \II + \tfrac{2}{3} \Delta \tau \DD \right) \ww^{k+1} \\
 = \left( \tfrac{4}{3} \ww^k - \tfrac{1}{3} \ww^{k-1} \right)
+ \Delta \tau \JJ \left( \tfrac{4}{3} \ww^k - \tfrac{2}{3} \ww^{k-1} \right)
+ \tfrac{2}{3} \Delta \tau \ff
\end{equation}
for European options and
\begin{equation}\label{BDF2step}
\begin{split}
& \left( \II + \tfrac{2}{3} \Delta \tau \DD + \tfrac{1}{\varepsilon}
\diag \left( \max \left\{ -\ww^{k+1},\, 0\right\} \right)
\right) \ww^{k+1} \\
&{} = \left( \tfrac{4}{3} \ww^k - \tfrac{1}{3} \ww^{k-1} \right)
+ \Delta \tau \JJ \left( \tfrac{4}{3} \ww^k - \tfrac{2}{3} \ww^{k-1} \right)
+ \tfrac{2}{3} \Delta \tau \ff
\end{split}
\end{equation}
for American options,
where the matrices $\JJ$ and $\DD$ corresponds to the terms due to the
jumps and the rest, respectively. The vector $\ff$ contains the grid
point values of $e^{-r \tau} (L + r) g$ and $Lg$. The operator
$\diag ( \cdot )$ gives a diagonal matrix with the diagonal entries
defined by the argument vector. The maximum is taken componentwise.
The systems \eqref{BDF2stepe} and \eqref{BDF2step} can be expressed
more compactly as
\begin{equation}
\AA \ww^{k+1}  = \rr^{k+1}
\end{equation}
and
\begin{equation}
\left( \AA + \tfrac{1}{\varepsilon}
\diag \left( \max \left\{ -\ww^{k+1},\, 0\right\} \right)
\right) \ww^{k+1}  = \rr^{k+1}
\end{equation}
with suitably defined $\AA$ and $\rr^{k+1}$.
The discrete counterpart of the Lagrange multiplier $\lambda$ in
\eqref{lambdadef} reads
\begin{equation}
\llambda^{k+1} = -\tfrac{1}{\varepsilon}
\diag \left( \max \left\{ -\ww^{k+1},\, 0\right\} \right) \ww^{k+1}.
\end{equation}

\section{Reduced Order Models}
\label{sect:ROMs}

Let $\UU \in {\mathbb R}^{N \times n}$ be the basis for $\ww$
with $n \ll N$. These basis are constructed by applying POD to
a collection of solution snapshots. A solution
snapshot, or simply a snapshot, is defined as a state vector $\ww^k$ computed as
the solution of~\eqref{BDF2step} for some instance of its parameters. A solution
matrix is defined as a matrix whose columns are individual snapshots.

To construct $\UU$, the following optimization
problem is solved
\begin{equation}\label{eq:SVDpb}
\underset{\UU\in\mathbb{R}^{N \times n},\,\VV \in\mathbb{R}^{n \times K}}
{\text{minimize}}
\displaystyle \| \XX - \UU \VV \|_F^2,
\end{equation}
where $K$ is the number of solution snapshots. Hence, the basis $\UU$ is
comprised of the first $n$ left singular vectors of the snapshot matrix $\XX$
and $\VV = \SS \WW^T$, where $\SS$
is the diagonal matrix of the first $n$ singular values of $\SS$, and $\WW$ is
the matrix of its first $n$ right singular vectors.

For European options the reduced solution $\ww = \UU \ww_r$
is governed by 
\begin{equation}\label{rom_e}
\UU^T \AA \UU \ww_r^{k+1} = \UU^T \rr^{k+1}.
\end{equation}
This ROM has the form
\begin{equation}\label{rom_fe}
\AA_r \ww_r^{k+1} = \rr_r^{k+1},
\end{equation}
where $\AA_r = \UU^T \AA \UU$ is precomputed offline while
the right-hand side $\rr_r^{k+1} = \UU^T \rr$ can be computed
efficiently online with the number of operations depending on $n$.
Thus, the online computational cost of forming and solving the
problems \eqref{rom_fe} scales with the size $n$ of the reduced
basis and it does not depend on the size $N$ of FOM.

For American options the reduced solution $\ww = \UU \ww_r$ is
governed by 
\begin{equation}\label{rom_1} 
 	\left( \UU^T \AA \UU + \tfrac{1}{\varepsilon} \UU^T \diag \left(
	\max \left\{ - \UU \ww_r^{k+1},\, 0\right\}
	\right) \UU \right)  \ww_r^{k+1} \\
	= \UU^T \rr^{k+1}.
\end{equation}
The product $\UU^T \diag \left(
\max \left\{ - \UU \ww_r^{k+1},\, 0\right\}
\right) \UU$ is the only product in~\eqref{rom_1} that cannot be precomputed
offline. Since the cost of evaluating this product scales with the size of the
full order model, Eq.~\eqref{rom_1} does not offer major computational savings.

To attain computational savings, the traditional approach involves including a
second layer of approximation, sometimes called ``hyper-reduction''. One of the
most popular hyper-reduction approaches is the Discrete Empirical Interpolation
Method (DEIM)~\cite{chaturantabut2010}. We recapitulate the traditional DEIM
algorithm as a starting point for our innovation. 

Let $\UU_\lambda \in {\mathbb R}^{N \times n_{\lambda}}$ be basis for $\max \left\{
- \UU \ww_r^{k+1},\, 0\right\}$, thus
\begin{equation}\label{assumption}
	\UU_\lambda \hh_r \approx \max \left\{ - \UU \ww_r^{k+1},\, 0\right\},
\end{equation}
where $\hh_r$ is the corresponding coefficient vector. The vector $\hh_r$ can be
determined by selecting $m$ unique rows from the overdetermined
system $\UU_\lambda \hh_r \approx \max \left\{ - \UU \ww_r^{k+1},\, 0\right\}$. 
Specifically, consider a binary matrix $\PP \in \{0,1\}^{N \times n_\lambda}$ satisfying
$\PP^T \PP = \II_{n_\lambda}$. Assuming
$\PP^T\UU$ is nonsingular, the coefficient vector $\hh_r$ can be determined
uniquely from
\begin{equation}
	\PP^T \max \left\{ - \UU \ww_r^{k+1},\, 0\right\} = (\PP^T\UU_{\lambda})\hh_r
\end{equation}
and the final approximation is 
\begin{align}
	\max \left\{ - \UU \ww_r^{k+1},\, 0\right\} \approx
		\UU_\lambda \hh_r &=
	 \UU_\lambda (\PP^T \UU_\lambda)^{-1} \PP^T \max \left\{ - \UU \ww_r^{k+1},\, 0\right\}\\
		&=\widetilde{\UU}_\lambda \max \left\{ - \CC \ww_r^{k+1},\,0\right\},
\end{align}
where $\widetilde{\UU}_\lambda = \UU_\lambda (\PP^T \UU_\lambda)^{-1}$, and 
$\CC = \PP^T \UU$.

Thus, the product 
$\UU^T \diag \left(\max \left\{ - \UU \ww_r^{k+1},\, 0\right\} \right) \UU$ in
Eq.~\eqref{rom_1}
is approximated by 
$\UU^T \diag \left(\widetilde{\UU}_\lambda \max \left\{ - \CC \ww_r^{k+1},\, 0\right\} \right) \UU$ 
that, unlike its predecessor, can be computed efficiently online. In
particular
\begin{equation}
	\UU^T \diag \left(\widetilde{\UU}_\lambda \max \left\{ - \CC \ww_r^{k+1},\, 0\right\} \right) \UU
	=
	\sum_{i=1}^{n_\lambda} \KK_i \max \left\{ -\left[\CC\ww_r^{k+1}\right]_i,\, 0 \right\},
\end{equation}
where $\KK_i = \UU^T \diag \left( \left[\widetilde{\UU}_\lambda\right]_{:,i} \right) \UU$ and
$\left[\widetilde{\UU}_\lambda\right]_{:,i}$
refers to the $i^{\text{th}}$-column of $\widetilde{\UU}_\lambda$. The matrices
$\KK_i$ can be computed offline, once and for all, while 
$\max \left\{ -\left[\CC \ww_r^{k+1}\right]_i,\, 0 \right\}$ 
can be computed efficiently online
since $\CC \in \mathbb{R}^{n_\lambda \times n}$ does not scale with
the size of the full order model. 

Although this straightforward implementation of DEIM succeeds in reducing the
computational complexity of the ROM, this approach cannot be expected to yield accurate
price predictions because DEIM does not enforce non-negativity.  
Even if the basis $\UU_\lambda$ are constructed to be non-negative a priori, using, for
example, NNMF, non-negativity is still not guaranteed because
$\widetilde{\UU}_\lambda = \UU_\lambda (\PP^T \UU_\lambda)^{-1}$ 
is not guaranteed to be non-negative. One possible remedy is use instead a
non-negative
variation of the DEIM, called NNDEIM~\cite{amsallem2016}. Yet another remedy
involves an angle-greedy procedure for constructing the non-negative
bases~\cite{Burkovska15}. In this work,
we introduce an alternative approach that does not require computation of
non-negative basis for the Lagrange multipliers. 

Our proposed approach is based on the fact that accurate price predictions do
not necessarily require accurate approximations of the Lagrange multipliers. 
In particular, requiring that $\UU_\lambda \hh_r \approx \max \left\{-\UU\ww_r^{k+1},0
\right\}$ may not be necessary. This has been observed in practice for the
reduction of structural contact problems~\cite{Balajewicz15}. Our numerical
experiments summarized in this paper suggest that using the binary matrix $\PP$
as the basis for the Lagrange multipliers performs remarkably well for all
reproductive and predictive simulations considered. With
this approximation, the reduced order model simplifies considerably. In
particular, with $\UU_\lambda = \PP$, $\widetilde{\UU}_\lambda = \PP$ and
thus, the product
$\UU^T \diag \left(\max \left\{ - \UU \ww_r^{k+1},\, 0\right\} \right) \UU$ in
Eq.~\eqref{rom_1} is approximated by the relatively simple product 
$\CC^T \diag \left(\max \left\{ - \CC \ww_r^{k+1},\, 0\right\} \right) \CC$. 
Thus, the final form of the ROM is as follows 
\begin{equation}\label{rom_2} 
 \left( \AA_r 
	 + \tfrac{1}{\varepsilon} \CC^T \diag \left( \max \left\{ - \CC \ww_r^{k+1},\,
 0\right\} \right) \CC \right)  \ww_r^{k+1} \\
= \rr^{k+1}_r,
\end{equation}
where $\AA_r = \UU^T \AA \UU$, and $\rr_r = \UU^T \rr$.
All components in Equation~\eqref{rom_2} scale with the size of the
reduced order model. 

Finally, to construct the selection matrix $\PP$, the standard DEIM algorithm
for selecting the interpolation indices is utilized~\cite{chaturantabut2010}.
However, in our proposed approach, the DEIM algorithm is applied to $\mm\odot\UU_{:,i}$, for $i
=1,2,\ldots,n$, where $\mm \in \{0,1\}^{N \times 1}$ is a binary mask vector.
The non-zero elements of the mask vector $\mm$, correspond to elements in the
snapshots where early exercise has occurred at least once, that is, elements $j$
such that $\UU_{j,i} \le 0$ for any $i$. The binary mask vector ensures
consistency with the nonlinear function that is being approximated, i.e. $\max
\left\{ - \UU \ww_r^{k+1},\, 0\right\}$. 

\section{Numerical Experiments}
\label{sect:Numerics}
All numerical examples considered here price a European put option and an American
put option with the strike price $K = 100$ and the expiry $T = 0.5$. Only at the
money options are considered, that is, the value of $u$ at $s = K$ is sought.
Under the stochastic volatility models the value of $u$ is computed at the
instantaneous variance $v = \theta$. The full order models are discretized using quadratically
refined spatial grids similar to ones employed by the FD-NU method in
\cite{vonSydow15b}. The $s$-grid is defined by
$s_i = \left[ \left(\tfrac{i}{\alpha N_s} - 1 \right)
\left|\tfrac{i}{\alpha N_s} - 1 \right| + 1 \right] K$, $i = 0, 1, \ldots, N_s$
with $\alpha = \tfrac{3}{8}$. For the stochastic volatility models
the variance grid is defined by
$v_j = \left(\tfrac{j}{N_v}\right)^2 v_{\max}$ with $v_{\max} = 1$.
The uniform time steps are given by $\Delta \tau = \tfrac{1}{N_\tau} T$.
In the experiments the number of spatial and temporal steps are chosen
to be $N_s = 128$, $N_v = 64$, and $N_\tau = 32$. With this choice and
the employed parameter ranges the absolute discretization error is about
$10^{-2}$ or less. In the case of the American option, an iteration
reduces the penalty parameter $\varepsilon$ with the five values $10^{-1}$,
$10^{-2}$, $10^{-3}$, $10^{-4}$, and $10^{-5}$. This is the main
reason for higher run times with the American option.

The snapshot matrix $\XX$ is given by all vectors $\ww^k$,
$k = 1, 2, \ldots, N_\tau$, in all training runs.
For these training runs each model parameter is sampled at its
extreme values and at the midpoint between them. Thus, with two,
five, and eight model parameters there are $3^2 = 9$, $3^5 = 243$,
and $3^8 = 6561$ training runs, respectively. In the predictive
ROM simulations, each parameter has two values which are the midpoint
values between the values used in the training. Thus, with two,
five, and eight model parameters there are $2^2 = 4$, $3^5 = 32$,
and $2^8 = 256$ prediction runs, respectively. The sizes of
the two reduced basis given by $n$ and $n_{\lambda}$ are chosen to
be the same. The measured error is the absolute difference between
the prices given by the reduced order model and the full order model.

All errors shown in Figures~\ref{fig:BlackScholes}--~\ref{fig:Bates} are
computed for the predictive simulations. That is, for simulations with
parameters not included in the training simulations used to generate the ROMs. 

\subsection{Black--Scholes Model}
\begin{figure}[tb]
	\begin{centering}
	\includegraphics[width=0.8\textwidth]{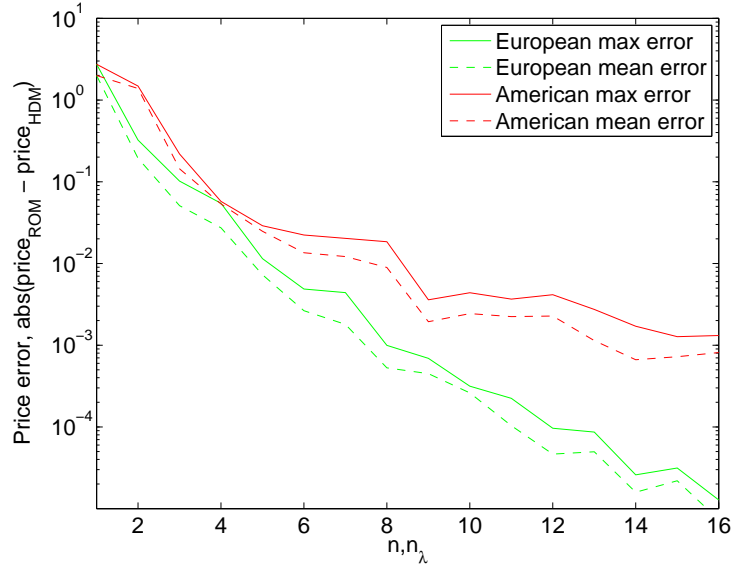}
	\caption{Under the Black--Scholes model the error with respect to the
                 the reduced basis size $n = n_{\lambda}$}
	\label{fig:BlackScholes}
	\end{centering}
\end{figure}

The model parameters for the Black--Scholes model are varied in the range:
\begin{equation}
(r,\, \sigma_s) \in [0.025,\, 0.035] \times [0.35,\, 0.45].
\end{equation}
The price of the European and American options vary roughly in
the ranges $[8.91, 11.94]$ and $[9.06, 12.04]$, respectively.
Figure \ref{fig:BlackScholes} shows the reduction of the maximum and
mean errors of the price of these options with the growth of
the reduced basis sizes $n = n_{\lambda}$.

\subsection{Merton Model}
\begin{figure}[tb]
	\begin{centering}
	\includegraphics[width=0.8\textwidth]{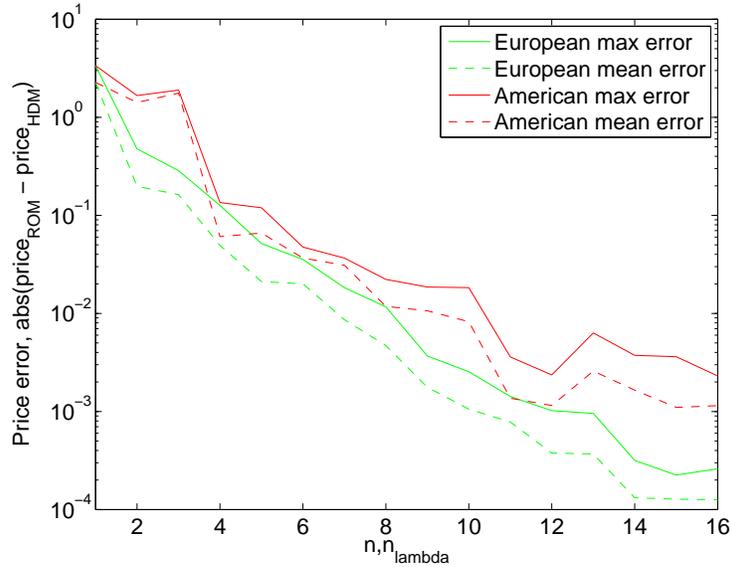}
	\caption{Under the Merton model the error with respect to the
                 the reduced basis size $n = n_{\lambda}$}
	\label{fig:Merton}
	\end{centering}
\end{figure}

The model parameters for the Merton model are varied in the range:
\begin{equation}
(r,\, \sigma_s,\, \mu,\, \delta,\, \gamma) \in
[0.025,\, 0.035] \times [0.35,\, 0.45] \times [0.15,\, 0.25] \times
[0.3,\, 0.5] \times [-0.7,\, -0.3].
\end{equation}
The price of the European and American options vary roughly in
the ranges $[9.50, 13.97]$ and $[9.65, 14.08]$, respectively.
Figure \ref{fig:Merton} shows the reduction of the maximum and
mean errors of the price of these options with the growth of
the reduced basis sizes $n = n_{\lambda}$.

\subsection{Heston Model}
\begin{figure}[tb]
	\begin{centering}
	\includegraphics[width=0.8\textwidth]{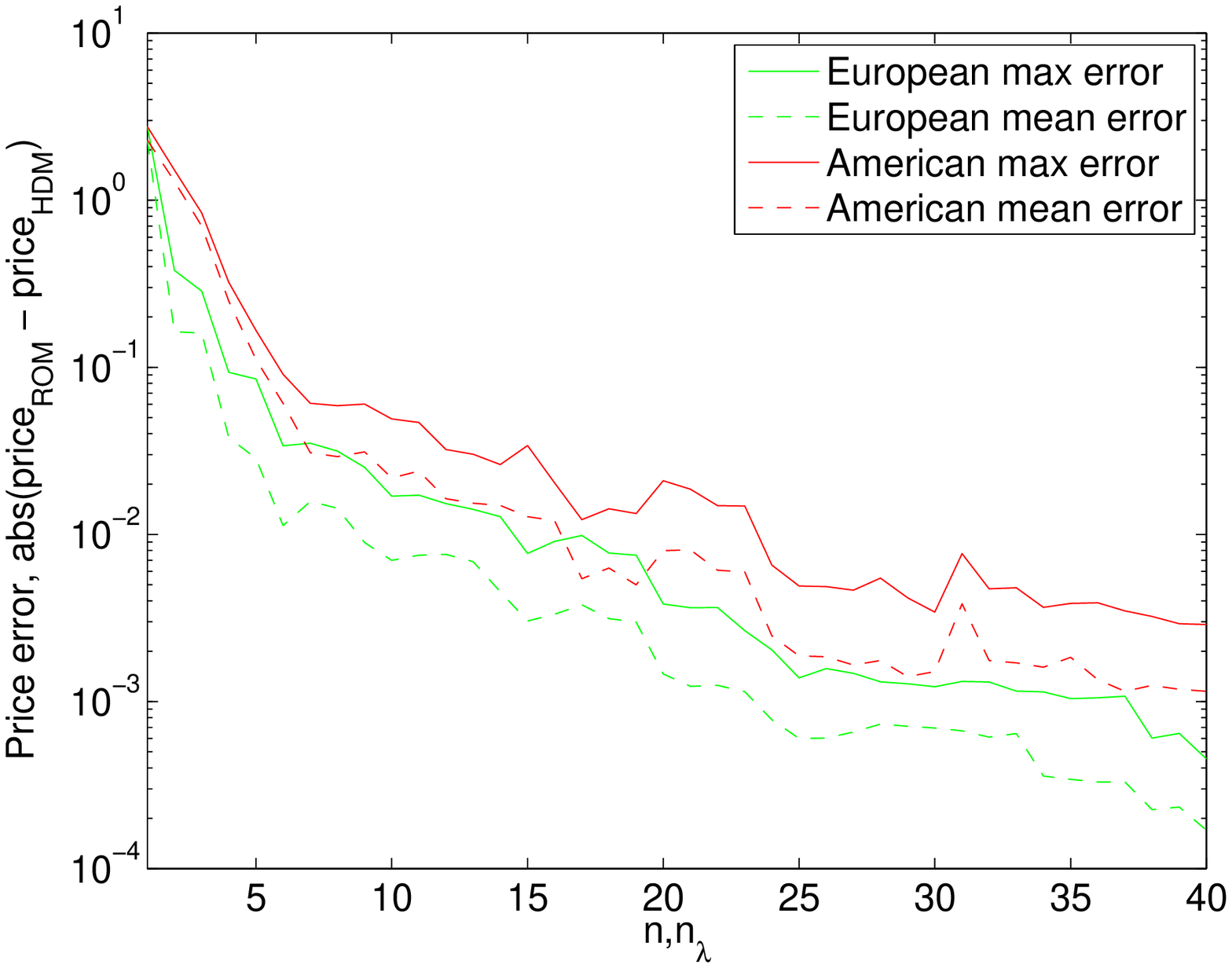}
	\caption{Under the Heston model the error with respect to the
                 the reduced basis size $n = n_{\lambda}$}
	\label{fig:Heston}
	\end{centering}
\end{figure}

The model parameters for the Heston model are varied in the range:
\begin{equation}
(r,\, \kappa,\, \theta,\, \sigma_v,\, \rho) \in
[0.025,\, 0.035] \times [3,\, 5] \times [0.35^2,\, 0.45^2] \times
[0.35,\, 0.45] \times [-0.75,\, -0.25].
\end{equation}
The price of the European and American options vary roughly in
the ranges $[8.72, 11.88]$ and $[8.87, 11.98]$, respectively.
Figure \ref{fig:Heston} shows the reduction of the maximum and
mean errors of the price of these options with the growth of
the reduced basis sizes $n = n_{\lambda}$.

\subsection{Bates Model}
\begin{figure}[tb]
	\begin{centering}
	\includegraphics[width=0.8\textwidth]{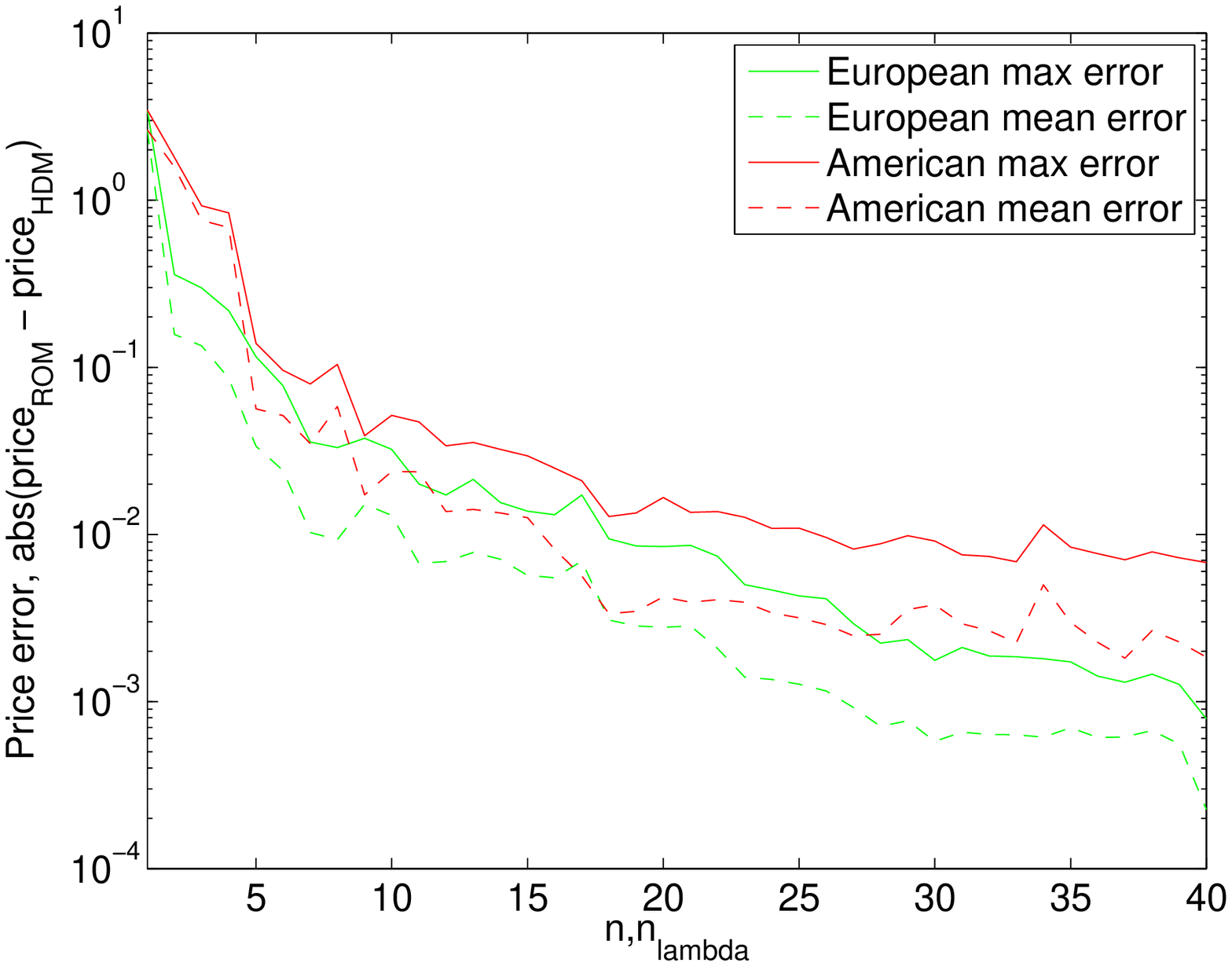}
	\caption{Under the Bates model the error with respect to the
                 the reduced basis size $n = n_{\lambda}$}
	\label{fig:Bates}
	\end{centering}
\end{figure}

The model parameters for the Bates model are varied in the range:
\begin{equation}
\begin{split}
(r,\, \kappa,\, \theta,\, \sigma_v,\, \rho,\, \mu,\, \delta,\, \gamma) \in {}
& [0.025,\, 0.035] \times [3,\, 5] \times
[0.35^2,\, 0.45^2] \times [0.35,\, 0.45] \times \\
& [-0.75,\, -0.25] \times [0.15,\, 0.25] \times
[0.3,\, 0.5] \times [-0.7,\, -0.3].
\end{split}
\end{equation}
The price of the European and American options vary roughly in
the ranges $[9.38, 13.95]$ and $[9.53, 14.07]$, respectively.
Figure \ref{fig:Bates} shows the reduction of the maximum and
mean errors the price of these options with the growth of
the reduced basis sizes $n = n_{\lambda}$. We note that for
this model essentially the same errors can be obtained based
only on $2^8 = 256$ training runs sampling the extreme values
of the model parameters.

\subsection{Computational Speed-up} 
For each problem considered, the speed-up factor delivered by
its ROM for the online computations is reported
in Table~\ref{tab:speedupe} for the European option and
in Table~\ref{tab:speedup} for the American option.
All models are solved in MATLAB on a Intel Xeon 2.6GHz CPU and all
CPU times were measured using the \verb=tic-toc= function on a single
computational thread via the \verb=-singleCompThread= start-up option.
A ROM is integrated in time using the same scheme and time-step used
to solve its corresponding FOM; see Section~\ref{sect:FOMs} for details.
The online speed-up is calculated by evaluating the ratio between
the time-integration of the FOM and the time-integration of the ROM. 

\begin{table}
\caption{For the European option CPU times in seconds for online computations.}
\centering
\begin{tabular}{l|c c|c c|c}
                & \multicolumn{2}{c|}{FOM} & \multicolumn{2}{c|}{ROM} \\
Model           & unknowns & CPU time & unknowns & CPU time & speed-up  \\
\hline
Black--Scholes  &\phantom{8}127 &  0.0011   &     16 &  0.00064 & 1.7 \\
Merton          &\phantom{8}127 &  0.0022   &     16 &  0.00084 & 2.6 \\
Heston          &   8255 &  0.16\phantom{99}&     40 &  0.0011\phantom{9} & 145 \\
Bates           &   8255 &  0.36\phantom{99}&     40 &  0.0015\phantom{9} & 240 \\
\end{tabular}
\label{tab:speedupe}
\end{table}

\begin{table}
\caption{For the American option CPU times in seconds for online computations.}
\centering
\begin{tabular}{l|c c|c c|c}
                & \multicolumn{2}{c|}{FOM} & \multicolumn{2}{c|}{ROM} \\
Model           & unknowns & CPU time & unknowns & CPU time & speed-up  \\
\hline
Black--Scholes  &\phantom{8}127 &  0.026   &     16 &  0.025   & 1.0 \\
Merton          &\phantom{8}127 &  0.027   &     16 &  0.026   & 1.0 \\
Heston          &   8255 &  7.9\phantom{99}&     40 &  0.034   & 232 \\
Bates           &   8255 &  8.0\phantom{99}&     40 &  0.034   & 235 \\
\end{tabular}
\label{tab:speedup}
\end{table}

\section{Conclusions}
\label{sec:conclusions}
Reduced order models (ROMs) were constructed for pricing European and American
options under jump-diffusion and stochastic volatility models.  For American
options they are based on a penalty formulation of the linear complementarity
problem. The finite difference discretized differential operator is projected
using basis resulting from a proper orthogonal decomposition. The grid points
for the penalty term are chosen using the discrete empirical interpolation
method.  In numerical experiments, from two to eight model parameters are varied
in a given range. For the one-dimensional Black--Scholes and Merton models about
16 ROM basis vectors were enough to reach 0.1\% accuracy for the considered
American option. For the European option about 8 basis vectors lead to this
accuracy. For the two-dimensional Heston and Bates models about 40 basis vectors
were needed to reach the same accuracy for the American option. Slightly less
basis vectors lead to the same accuracy for the European option.  For these
two-dimensional models the computational speed-up was over 200 when the full
order model (FOM) and ROM have roughly the same 0.1\% accuracy level for the
American option.  For the European option the solution of the FOM and the ROM
under the Bates model required about 0.36 and 0.0015 seconds, respectively.  For
the American option the solution of the FOM and the ROM for two-dimensional
models required about 8 and 0.034 seconds, respectively.  With the
one-dimensional models the speed-up was negligible.  Particularly the results
with the Bates model and eight parameters varying are impressive. For
one-dimensional models probably for most applications FOMs are sufficiently
fast. For two-dimensional models often FOMs are computationally too expensive
and in such cases the proposed ROMs can enable the use of these models.
Performance of the proposed ROM approach is quite similar to previous approaches
based on the NNMF. For example, the maximum ROM price error using 40 basis
vectors under the Heston model using the proposed approach and the previous
approach based on NNMF is $2.9 \times 10^{-3}$, and $4.2 \times 10^{-3}$,
respectively. While for the Bates model, the maximum ROM price error using 40
basis vectors using the proposed approach and the previous approach based on
NNMF is $6.8 \times 10^{-3}$, and $4.0 \times 10^{-3}$ respectively. A potential
application for these ROMs is the calibration of the model parameters based on
market data. With a least squares calibration formulation, option prices and
their sensitivities can be computed quickly and accurately for varying
parameters by employing ROMs.

%
\label{sect:bib}
\bibliography{main}

\end{document}